\documentclass[article,twocolumn]{revtex4-1} 

\usepackage{amsmath}  
\usepackage{amsfonts} 
\usepackage{graphicx} 
\usepackage{hyperref, bookmark}						
\hypersetup{colorlinks=true,linkcolor=black,citecolor=blue,urlcolor=blue}

\begin{document}
    
\title{Room-temperature solid-state masers as low-noise amplifiers to facilitate deep-space missions using small spacecraft}

\author{Carlos Barbero Rodr\'iguez}
\email{carlos.barbero19@imperial.ac.uk} 
\affiliation{(Imperial College London, Department of Materials, United Kingdom, SW7 2AZ)}

\date{September 25, 2023}

\begin{abstract}
An increasing number of small ventures are launching missions to space with small volume satellite platforms. These small spacecraft are now being seriously considered for deep-space missions, creating a need for ground stations capable of detecting the faint signals they will transmit to Earth. Here, recent developments in room-temperature solid-state masers are reviewed to determine their readiness for use as a cheap low-noise amplifier for deep-space communications. Masers based on Pentacene-doped Para-terphenyl (Pc:PTP), Pentacene-doped Picene, Diazapentacene-doped Para-Terphenyl (DAP:PTP), Phenazine/1,2,4,5-Tetracyanobenzene (PNZ/TCNB) co-crystal, NV Diamond, Cuprous Oxide, and Silicon Carbide are considered for comparison. Pc:PTP offers good spin polarisation density and output power but suffers from thermal dissipation problems, DAP:PTP may help to obtain a lower threshold power than that achieved with Pentacene, PNZ/TCNB stands out in spin polarization density but has not achieved room-temperature masing, and NV Diamond is the only medium to have sustained continuous operation but has very limited power output. The other gain media proposed offer theoretical advantages but have not been tested in a working maser device.
\end{abstract}

\maketitle 

\section{Introduction}\label{sec:introduction}
In the recent years, there has been a drastic reduction in the cost of launching a vehicle to space. This change has been accompanied by miniaturization of electronics. Now small companies and universities can afford to develop their own space missions. The CubeSat is a popular standard defined as a payload with 10x10x10cm dimensions, with many satellites based on it currently orbiting the Earth. A newer standard was proposed in 2009 for a “picosatellite” called the PocketCube with a 5x5x5cm size, making it approximately 1/8 of the price to launch with respect to a CubeSat. These picosatellites have an estimated launch cost to Low Earth Orbit (LEO) of \textdollar20k, making them very attractive for universities and crowd-funded projects. \citep{ElwoodAgasid2018}

Although these satellite platforms are currently mostly deployed in LEO, they are now being seriously considered for lunar missions and deep-space exploration. Examples of such missions are already underway, including two lunar CubeSat orbiters launched by NASA \citep{BosanacNatasha2018Tdfa,AgasidElwood2021CAPS} in 2022 and another two being studied by ESA\citep{CERVONE2022309,KruzeleckyRoman2022LVaM}. The age of deep-space small spacecraft seems to be close, but there are still technical problems to be tackled.

The focus of this paper is on the challenge of communication with small spacecraft, particularly of amplifying the faint signal coming from the spacecraft at the ground station. The different room-temperature solid-state masers under research will be presented, evaluating their viability as low-noise amplifiers for use in low-cost ground stations.
\section{Challenges of Deep-Space Signal Reception}\label{sec:challenges}
The nature of deep-space communications is highly asymmetric. The spacecraft has severely limited power, volume, and weight, resulting in poor signal amplification for transmission and reception. The same limitations are not present in the ground station. This means that high-power ground transmitters with large parabolic antennae are used in Earth to communicate with spacecraft carrying low-power transponders with poorly focused antennae.  A representative example of this is the Voyager 1 mission’s photographs of Jupiter, where a 64m diameter parabolic dish was used to receive data from a 20W transmitter with a 3.7m antenna located 680 million kilometres away from Earth \citep{EdelsonRE1979VTTB}. 

For space communications, the ratio of received power to transmitted power $\frac{P_r}{P_t}$ is given by the “space loss” \citep{alma991000261722601591}. It is obtained from the Friis Transmission Formula\citep{1697062}, where $A_t$ and $A_r$ are the effective areas of the transmitting and receiving antennae:
\begin{equation}
    \frac{P_r}{P_t} = A_rA_t\frac{1}{d^2\lambda^2}
\end{equation}

Here $d$ is the distance to the ground station and \( \lambda{} \) is the signal wavelength. The signal reaching Earth would be very faint for deep-space missions as the distances are in the order of millions of kilometres and the spacecraft antenna area is severely limited. An added complication is that space contains noise equivalent to 3K temperature, the cosmic microwave background, which will be added to the noise temperature of the receiving system \citep{EdelsonRE1979VTTB}.  The signal received at the ground station will be similar or lower than the noise power (see Appendix). The maximum achievable bit rate \textit{R} in bits per second for a communication channel can be obtained from Shannon’s law with the following equation \citep{alma991000261722601591}:

\begin{equation}
    R = B\text{log}_2(1+\text{SNR})
    \label{eq:ShannonLimit}
\end{equation}

Here $B$ is the bandwidth, with a typical value of 36MHz in commercial satellite systems\citep{alma991000261722601591}, and SNR is the signal to noise power ratio. There exists a lower SNR value for reliable communication given by the threshold \(\frac{E_b}{N_0}>-1.6\text{dB}\), where \(E_b\) is the energy per bit of information and \(N_0\) is the noise power spectral density \citep{madhow_2008}. Hence, without the ability to increase transmitted power, the noise of the receiving system must be minimized. In this application masers have found a historical use due to their very low noise level and high gain, which renders noise further in the receiving circuit negligible \citep{TaborW.J.1963MftT}. Masers were first used for space communications by NASA in the TELSTAR satellite\citep{TaborW.J.1963MftT} and continued their use in other missions such as Voyager\citep{EdelsonRE1979VTTB} and the TDRSS program\citep{QuinnR.B.1988A2ma}.  Although the technology has been successfully applied in space communications, it is reasonable to expect that the increased volume of low-budget missions from universities and small companies will require more communication capacity than that offered by the big players like NASA. This means that there is a need for cheaper and simpler set-ups for small ventures to afford building their own ground stations.

\section{Solid-State Masers as Low-Noise Amplifiers}\label{sec:maserPrinciples}
Maser (Microwave Amplification by Stimulated Emission of Radiation) operation relies on stimulated emission whereby a system relaxes to a low energy level triggered by an incident photon of energy equal to the energy difference between levels. This transition causes the emission of a coherent photon of equal frequency. If the rate of stimulated emission is higher than the rate of absorption, radiation travelling through the medium will be amplified by it, conserving its frequency and phase \citep{siegman}.

To achieve a higher rate of stimulated emission and net amplification, a large population inversion must be achieved where the higher energy level has larger electron population than the lower energy level. This is achieved in solid state masers with the three-level scheme proposed independently by Basov and Prokhorov in 1955, and by Bloembergen in 1956. \citep{orton}

\begin{figure}[h!]
  \centering
  \includegraphics[width=0.8\linewidth]{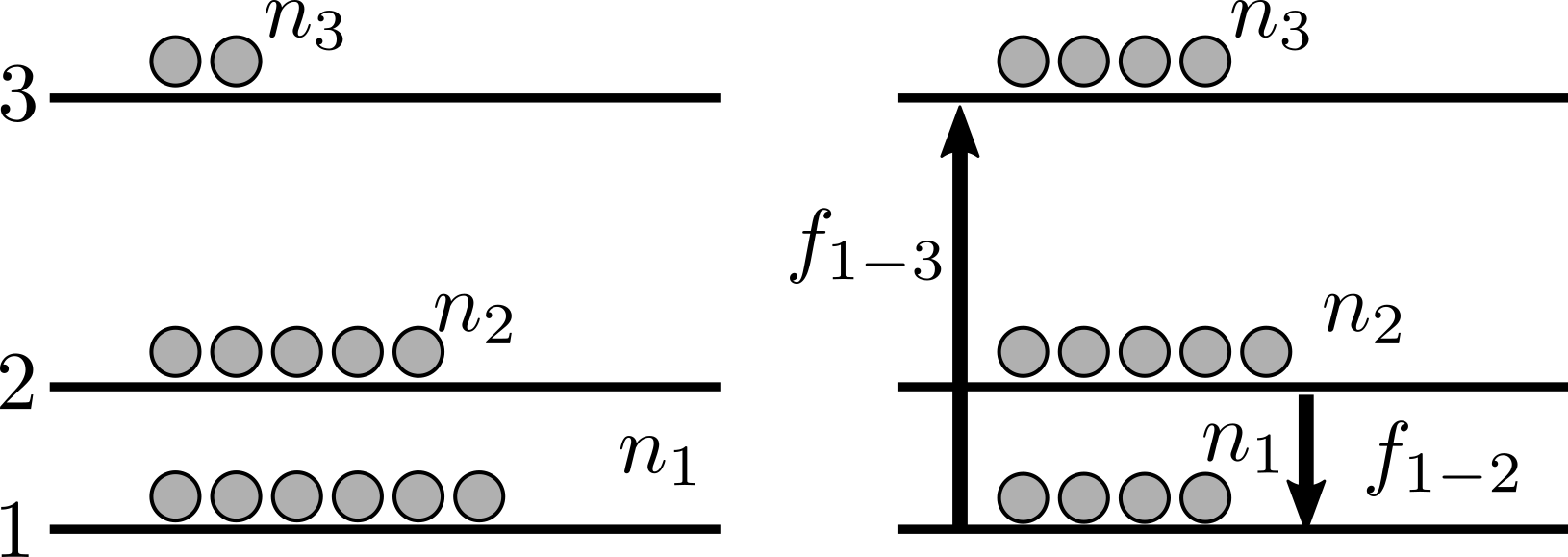}
  \caption{Illustration of the basic 3-level scheme before and after pumping. Populations and relevant transition frequencies labelled.}
  \label{fig:threelevel}
\end{figure}
A system is chosen with 3 states of ascending energy level 1,2 and 3 where their populations are expressed as $n_1$,$n_2$ and $n_3$ (Fig \ref{fig:threelevel}). The medium is exposed to radiation of energy equal to the difference between level 1 and level 3 at a frequency referred to as the pump frequency \(f_{1-3}\). Higher intensity pumping results in more transitions to the upper level. When both levels have the same population, the 1-3 transition is said to have been driven to saturation. Sufficient pumping power leads to population inversion between levels 2 and 1, making the material suitable for masing at a signal frequency of \(f_{1-2}\).\citep{orton}\citep{siegman}

The power gain of the maser when acting as an amplifier in a cavity is proportional to the inverse of the magnetic Q which depends on the following factors (As described in detail in Chapter 6 of reference \citep{siegman}):

\begin{equation}
\frac{1}{Q_m} \propto \frac{h f_{1-2}}{k_BT}\frac{N}{n}\frac{I\eta}{\Delta{}f_L}
\end{equation}

Where \textit{h} is Planck's constant, $k_B$ is the Boltzmann constant, \textit{T} is temperature in kelvin, $\eta$ is the magnetic fill factor, \textit{I} is the population inversion fraction, $\Delta{}f_L$ is the magnetic resonance linewidth,  \textit{N} is the density of polarized states and \textit{n} is the number of energy states in the system (3 for the theoretical Bloembergen scheme).

The available range of frequencies is determined by international communication guidelines so not much can be done about it \citep{alma991000261722601591}. The magnetic fill factor $\eta$ can be maximized with an appropriate cavity design with a small mode volume. The population inversion \textit{I} can be improved by using a higher frequency pump \citep{alma991000618553101591}, with optical pumping being chosen for room-temperature maser applications. The higher pump frequency comes at a cost as the energy efficiency of the maser is decreased \citep{orton}. This decrease in energy efficiency does not pose a problem in a practical amplifier though as ground stations don’t have a limited power budget. To maintain a strong inversion, longer relaxation times are desired for the 1-2 transition. The spin polarisation density \textit{N} needs to be maximized by having a high density of active species and the highest yield possible of the polarized states on pumping.

Conventional solid-state masers achieve the three-level system with a signal frequency in the microwave region using paramagnetic ions inside a crystal. The most used solid-state masers use Chromium ions in Ruby as a gain medium \citep{orton}, exploiting the Zeeman levels formed in the presence of a magnetic field \citep{siegman}. These masers achieve extremely low noise temperature, they can be frequency tuned by an external magnetic field, and their transition energy range can be broadened with an inhomogeneous magnetic field for use in space communications \citep{TaborW.J.1963MftT}. In order to sustain masing with a high enough gain, they are operated at cryogenic temperatures in the liquid helium range making them expensive to maintain. Another problem in their operation lies in the need for external electromagnets to cause Zeeman Splitting \citep{TaborW.J.1963MftT,siegman}. These requirements make conventional solid-state masers expensive to operate for prolonged periods. For these amplifiers to become widespread, a room temperature zero-field high gain maser is desired.

\section{Current Developments in Room-Temperature Solid-State Masers}
Current research into room-temperature solid-state masers can be broadly divided in two different branches. One is based on exploiting the non-degenerate triplet state splitting in organic molecules with low symmetry at zero magnetic field \citep{OXBORROWMark2012Rsm,BreezeJonathan2015EmPe,alma991000404984801591,https://doi.org/10.48550/arxiv.2211.06176,MoroFabrizio2022Rodm,NgWern2021EtTS}, and the other branch focuses on inorganic crystals \citep{BreezeJonathanD2018Crdm,Ziemkiewicz:18,Ziemkiewicz:19,Gottscholl_2022}. The recent advances in these two branches will be described with a focus on the viability for a high gain low noise amplifier.

\subsection{Organic Masers}

\begin{figure}
  \centering
  \includegraphics[width=0.9\linewidth]{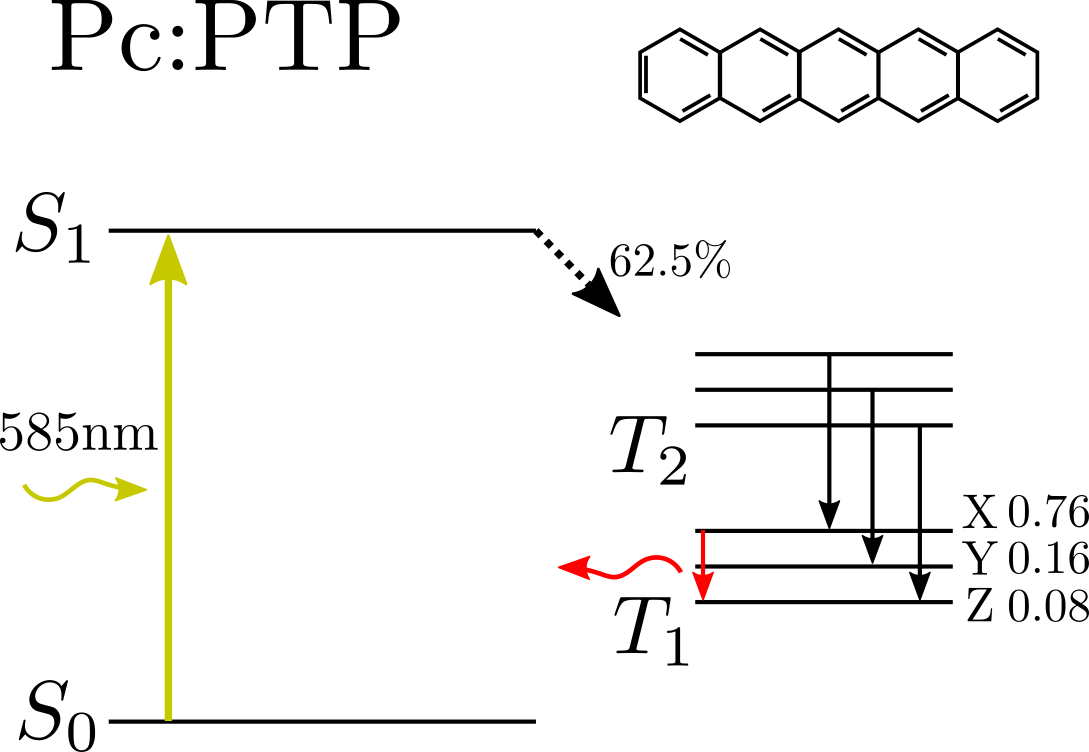}
  \caption{Jablonski diagram of the energy states of Pentacene during operation of the Pc:PTP maser. Quantum yield of intersystem crossing and population ratios are indicated on the diagram.}
  \label{fig:PcJablonski}
\end{figure}

In 2012, Oxborrow \textit{et al.}\citep{OXBORROWMark2012Rsm} created the first room-temperature solid state maser by using Pentacene-doped Para-terphenyl (Pc:PTP) as a gain medium. The crystal is placed in a sapphire ring inside a tuneable microwave cavity. The maser is optically pumped at 585nm wavelength by a pulsed laser to excite the Pentacene molecules from the ground singlet state to the excited singlet $S_1$. They then preferentially undergo intersystem crossing to the excited triplet state $T_2$ (yield of 62.5\%) and relax into the lower triplet state $T_1$. The triplet states of Pentacene exhibit zero-field splitting into 3 non-degenerate energy levels X, Y and Z in order of decreasing energy, with intersystem crossing populating them preferentially in the proportions 0.76:0.16:0.08 respectively. The population inversion is then exploited for masing by tuning the microwave cavity to the X-Z transition frequency of 1.45GHz \citep{OXBORROWMark2012Rsm,BreezeJonathan2015EmPe}. In Fig \ref{fig:PcJablonski}, a Jablonski diagram shows the relevant states and the population ratios of $T_1$. A threshold pump power of 230W was required and its operation was limited to pulsed mode mainly due to the heat build-up in the crystal and its low melting temperature \citep{OXBORROWMark2012Rsm}.

\begin{figure}
  \centering
  \includegraphics[width=0.8\linewidth]{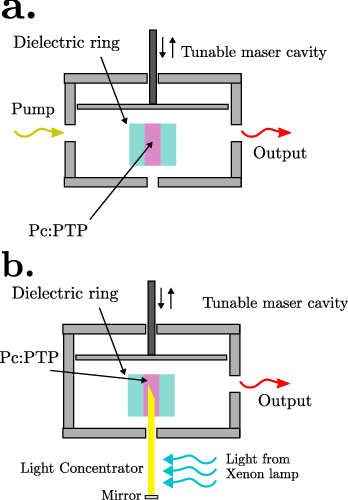}
  \caption{Simplified schematics of the Pc:PTP maser. \textbf{a.} shows the pumping scheme used in the original maser and the 2014 STO ring maser. \textbf{b.} shows the invasive pumping scheme with which quasi-continuous operation was achieved.}
  \label{fig:MaserAnatomy}s
\end{figure}

\begin{figure}
      \centering
      \includegraphics[width=\linewidth]{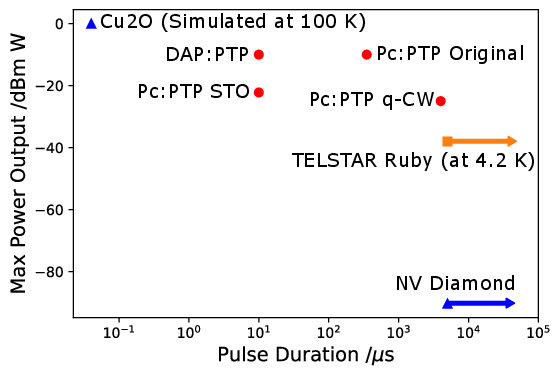}
      \caption{Maximum measured output power and pulse duration for different room-temperature solid-state masers. The red circles represent organic masers, the blue triangles are inorganic masers, and the orange square is the conventional maser used for communication with the NASA TELSTAR satellite. The two datapoints with an arrow pointing to the right indicate that continuous operation was achieved (effectively pulse duration $\rightarrow\infty$). The values given for $Cu_2O$ are based on several computer simulations for performance at 100K.}
      \label{fig:PowerTime}
\end{figure}

\begin{figure}
      \centering
      \includegraphics[width=\linewidth]{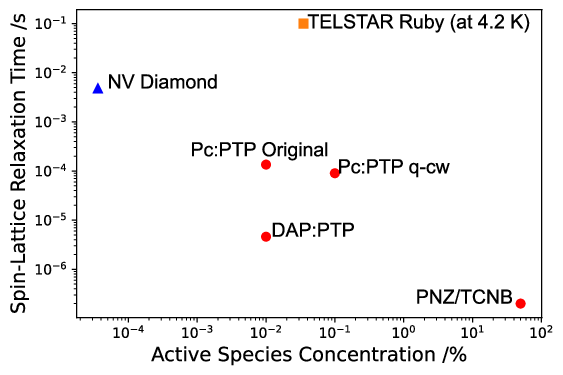}
      \caption{Spin-lattice relaxation time and active species concentration for different room-temperature solid-state masers. Notice the logarithmic scale on both axes. Red circles represent organic masers, blue triangles are inorganic masers, and the orange square is the conventional maser used for communication with the NASA TELSTAR satellite. The distribution of datapoints shows that the density and spin stability characteristics of the TELSTAR maser have not been matched in room temperature operation as no gain medium has been found with both high active species concentration and long room-temperature spin decay time.}
      \label{fig:DecayConc}
\end{figure}

The problem of overheating in the original Pentacene maser could be tackled by reducing the threshold pump power required for masing. Two factors needed to be optimized: the magnetic fill factor and the cavity losses. This was achieved in 2014 by taking advantage of the high permittivity of a Strontium Titanate (STO) ring to make a low mode volume, high Q factor maser cavity \citep{BreezeJonathan2015EmPe}. The result was a decrease in two orders of magnitude for the threshold pump power in the Pc:PTP maser. Further progress was made, achieving quasi continuous-wave (q-cw) operation for a duration of 4ms with a Pc:PTP maser \citep{alma991000404984801591} (See Fig \ref{fig:PowerTime} for comparison of maser pulse times). Using the STO cavity described, thermal issues in the maser were improved further by changing the pumping arrangement (See Fig \ref{fig:MaserAnatomy}). Instead of a pulsed laser incident on the Pc:PTP crystal, a xenon lamp was used with a wedge-shaped light concentrator which was driven inside the crystal (Invasive optical pumping). Having the pump light driven directly inside the crystal resulted in more uniform coverage of the gain medium and in a better interface, reducing heat losses and damage to the crystal. Despite these improvements, the heat dissipation problem still prevented continuous operation of the pentacene maser, which would be required for operating a low noise amplifier \citep{alma991000404984801591}. Overall, with an achieved doping concentration of 0.1\%\citep{alma991000404984801591} and a yield of 62.5\%, Pc:PTP has a high spin polarisation density (Fig \ref{fig:DecayConc}) and remains a very attractive medium for a high gain maser if the continuous operation issues are resolved.

Apart from the described improvements on the Pc:PTP maser, research has been conducted into alternative organic masing materials. As an alternative host to Para-terphenyl, Picene has been proposed as it exhibits better energy transfer to the triplet states in Pentacene \citep{MoroFabrizio2022Rodm}. A pulsed mode maser has also been built with Diazapentacene-doped Para-terphenyl (DAP:PTP), a desirable substitute to Pentacene which is capable of achieving a higher triplet yield of 67\%. DAP:PTP could be used to attain a lower pumping threshold power \citep{https://doi.org/10.48550/arxiv.2211.06176}. A new promising approach that has been studied is the use of organic co-crystals where much higher density of states could be achieved than what is possible with doping a host crystal (Fig \ref{fig:DecayConc}). Masing was tested for the Phenazine/1,2,4,5-Tetracyanobenzene co-crystal (PNZ/TCNB)\citep{NgWern2021EtTS} but was not achieved without cavity Q factor boosting by external amplifiers. It was determined in that same paper that a co-crystal needs to be found with longer spin relaxation times.


\subsection{Inorganic Masers}

\begin{figure}
  \centering
  \includegraphics[width=\linewidth]{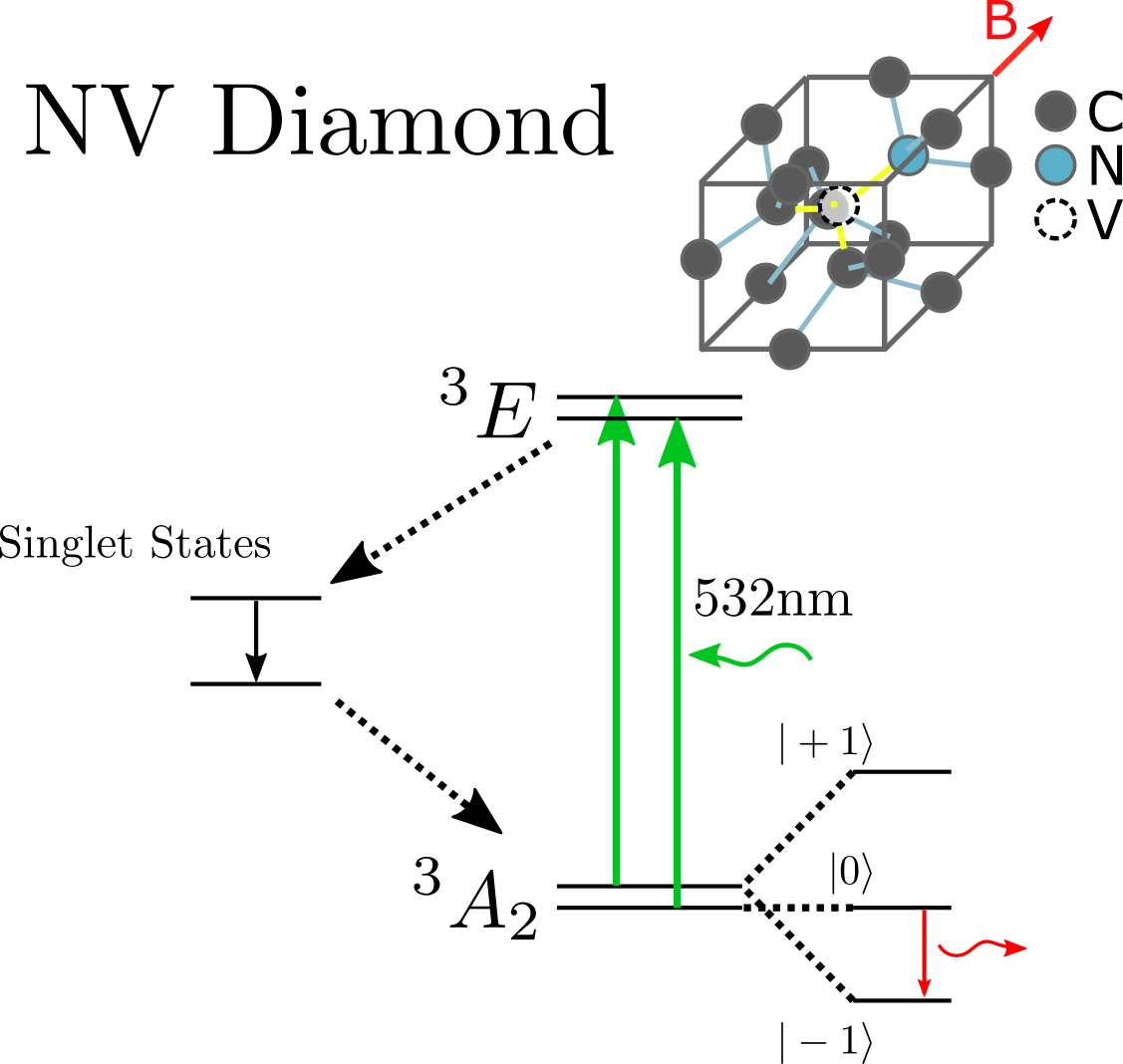}
  \caption{Jablonski diagram of the energy states of NV centers during operation of the NV diamond maser. Ground and excited triplet states labelled as $^3A_2$ and $^3E$, respectively. Splitting due to the presence of a magnetic field (B field) shown for $^3A_2$. Illustration of the NV-centre structure with direction of the magnetic field shown on top right corner.}
  \label{fig:NVJablonski}
\end{figure}

The representative technology for inorganic room-temperature solid-state masers is the NV diamond maser \citep{BreezeJonathanD2018Crdm}. It is based on paramagnetic spin levels in nitrogen-vacancy centres inside the diamond lattice (Fig \ref{fig:NVJablonski}). The triplet states in NV diamond have 3 sublevels: $|0\rangle$ $|+1\rangle$ and $|-1\rangle$, with the $|\pm1\rangle$ states being degenerate at zero field. To achieve masing, an externally applied magnetic field is required to separate the $|\pm1\rangle$ states and place the $|-1\rangle$ state at a lower energy level than the $|0\rangle$ state. The population inversion is achieved by using a 532nm optical pump to promote the system from the ground triplet to the excited triplet. The NV centre then undergoes intersystem crossing to the excited singlet and back to the ground triplet, where spin selective crossing causes a population inversion between $|0\rangle$ and $|-1\rangle$ as seen in Fig \ref{fig:NVJablonski}. The NV centres in diamond form long lived triplet states, making them suitable for room temperature operation. However, only low concentrations of NV centres could be introduced in diamond (Fig \ref{fig:DecayConc}), limiting the gain and output power. An added advantage of Diamond is its very high thermal conductivity and good mechanical properties, enabling continuous operation and making it more viable for low noise amplifiers. The requirement of an externally applied magnetic field also adds to the cost and complexity of such a system. However, a non-uniform magnetic field allows for frequency tuning and broadening of bandwidth as performed in the TELSTAR satellite communications \citep{TaborW.J.1963MftT}, which would benefit the maximum bit-rate obtainable.

Another recent proposal consists of exploiting the discrete energy levels of excitons in cuprous oxide (Cu$_2$O) to form a 3-level maser system \citep{Ziemkiewicz:18}. Excitons in semiconductors form when an electron is promoted to the conduction band and remains binded to a hole. The hole-electron system behaves similarly to a hydrogen atom and has discrete energy levels which can have transitions in the microwave range. These excitons have long relaxation times and can be tuned by an applied electric field, making it possible to create the three-level system with them. It has been claimed that the Cu$_2$O gain material can mase at room temperature and its operation at pulsed mode has been simulated computationally \citep{Ziemkiewicz:19} for an operating temperature of 100 K. Although the tunability by electric field is advantageous, the spin polarisation density is limited below that of NV diamond by Rydberg blockade and it hasn’t been proven to work at room temperature yet \citep{Ziemkiewicz:19}.

A promising gain medium in terms of scalability for mass production is that of spin defects in Silicon Carbide (SiC) located at Silicon vacancies. Gottscholl et al. \citep{Gottscholl_2022} have determined it to be a viable gain medium for room temperature masing although it has short spin decoherence times and requires a carefully aligned external magnetic field for operation, adding to the complexity of the set up. The doping concentration of vacancies may also be a limiting factor as with the NV diamond. Despite its disadvantages, this gain medium is promising as SiC is already manufactured in many forms for high power electronics \citep{Gottscholl_2022}, making it ready for mass production.

\section{Conclusion}
As it stands, there are no clear winners; here are the details. The Pc:PTP maser has undergone several improvements since its introduction in 2012 and shows promising results, but true continuous operation must be achieved. The NV diamond maser, with its good thermal and mechanical properties, is capable of continuous operation, but its performance as an amplifier is severely limited by the low NV centre concentration obtained. To better evaluate the suitability of these masers as amplifiers, their noise levels should be measured and compared to conventional masers. Alternative organic masing media have been proposed by using Picene as a host, using Diazapentacene instead of Pentancene, and by using a PNZ/TCNB co-crystal. Diazapentancene has been shown to improve performance while Picene has not been experimentally tested in a maser device and PNZ/TCNB has not achieved masing despite the high spin polarisation density it offers.  Inorganic masers with $SiC$ and $Cu_2O$ gain mediums have not yet been built but they offer the promise of ease of mass manufacture and tuning by electric fields respectively. All in all, room-temperature solid-state masers may soon make it possible to have a cheap low-noise amplifier set-up for space communications.

\begin{acknowledgments}

This paper was made possible by the feedback and explanations of Professor Mark Oxborrow.

\end{acknowledgments}

\section*{Author Declarations}
\subsection*{Conflicts of Interests}
    The author has no conflicts of interests to disclose.
    

\appendix*   

\section{Viability of use in space exploration}
Using the Friis Transmission Formula, some preliminary calculations may be performed to understand the problem of communications with small spacecraft. Here we assume a spacecraft equipped with a 1 Watt transmitter and a parabolic antenna of 0.0047$m^2$ effective area (roughly 10 centimetres in diameter). These specifications would correspond to a small cubesat like those seen in low earth orbit. The receiving antenna is assumed to be 30$m^2$ in effective area (roughly 8 metres in diameter), and the signal is at a frequency of 1.45GHz over a bandwidth of 1MHz. To calculate the SNR, a receiving system noise temperature of 50 Kelvin is used like that initially projected for the TELSTAR program \citep{TaborW.J.1963MftT}. The average distance from Earth to different celestial bodies is used\cite{NASADist}, but it should be noted that orbital trajectories of planets result in large variations in distance from the average. The power received is given as equivalent temperature for easier comparison with the noise level.

\begin{table}[h!]
\centering
\caption{Bit-rate and viability of communication for several astronomical distances.}
\begin{ruledtabular}
\begin{tabular}{p{1.5cm} p{1cm} p{2cm} p{1cm} c}
Average Distance (10$^6$km) & Received Power (dBm) & Received equivalent temp. (K) & Shannon Limit (Kbits/s)& $E_b/N_0$\\
\hline	
0.384 (Moon) & -136 & 1 600 000 & 15 000 & 33.4 \\
 228 (Mars) & -191 & 4.6 & 127 & -1.40 \\
 1400 (Saturn) & -207 & 0.00002 & 4 & -1.59 \\
\end{tabular}
\end{ruledtabular}
\label{tableAppendix}
\end{table}

From the results obtained, there is an evident loss of signal quality when exploring further planets: a 500x500 pixel 8-bit gray-scale image would take 0.1 seconds to send from the Moon and 500 seconds to send from Saturn when operating at Shannon's limit. Technical difficulties aside, it would not be advisable to operate at Shannon's limit from Saturn as the $E_b/N_0$ is close to the threshold for reliable communication. Another important observation is that in the cases of Mars and Saturn, the SNR is very small. From Equation \ref{eq:ShannonLimit}, when the SNR is very small, the increase in bitrate is linear with SNR, meaning that improvements in receiver noise level will have a noticeable impact. The low-noise requirements of this application justifies the use of a maser as an amplifier.

\bibliography{Refs}

\begin{thebibliography}{25}%
\makeatletter
\providecommand \@ifxundefined [1]{%
 \@ifx{#1\undefined}
}%
\providecommand \@ifnum [1]{%
 \ifnum #1\expandafter \@firstoftwo
 \else \expandafter \@secondoftwo
 \fi
}%
\providecommand \@ifx [1]{%
 \ifx #1\expandafter \@firstoftwo
 \else \expandafter \@secondoftwo
 \fi
}%
\providecommand \natexlab [1]{#1}%
\providecommand \enquote  [1]{``#1''}%
\providecommand \bibnamefont  [1]{#1}%
\providecommand \bibfnamefont [1]{#1}%
\providecommand \citenamefont [1]{#1}%
\providecommand \href@noop [0]{\@secondoftwo}%
\providecommand \href [0]{\begingroup \@sanitize@url \@href}%
\providecommand \@href[1]{\@@startlink{#1}\@@href}%
\providecommand \@@href[1]{\endgroup#1\@@endlink}%
\providecommand \@sanitize@url [0]{\catcode `\\12\catcode `\$12\catcode
  `\&12\catcode `\#12\catcode `\^12\catcode `\_12\catcode `\%12\relax}%
\providecommand \@@startlink[1]{}%
\providecommand \@@endlink[0]{}%
\providecommand \url  [0]{\begingroup\@sanitize@url \@url }%
\providecommand \@url [1]{\endgroup\@href {#1}{\urlprefix }}%
\providecommand \urlprefix  [0]{URL }%
\providecommand \Eprint [0]{\href }%
\providecommand \doibase [0]{http://dx.doi.org/}%
\providecommand \selectlanguage [0]{\@gobble}%
\providecommand \bibinfo  [0]{\@secondoftwo}%
\providecommand \bibfield  [0]{\@secondoftwo}%
\providecommand \translation [1]{[#1]}%
\providecommand \BibitemOpen [0]{}%
\providecommand \bibitemStop [0]{}%
\providecommand \bibitemNoStop [0]{.\EOS\space}%
\providecommand \EOS [0]{\spacefactor3000\relax}%
\providecommand \BibitemShut  [1]{\csname bibitem#1\endcsname}%
\let\auto@bib@innerbib\@empty
\bibitem [{\citenamefont {{Elwood Agasid}}\ \emph {et~al.}(2018)\citenamefont
  {{Elwood Agasid}}, \citenamefont {{Roland Burton}}, \citenamefont {{Roberto
  Carlino}}, \citenamefont {{Gregory Defouw}}, \citenamefont {{Andres Dono
  Perez}}, \citenamefont {{Arif G{\"{o}}ktuğ Karacalıoğlu}}, \citenamefont
  {{Bemjamin Klamm}}, \citenamefont {{Abraham Rademacher}}, \citenamefont
  {{James Schalkwyck}}, \citenamefont {{Rogan Shimmin}}, \citenamefont {{Julia
  Tilles}},\ and\ \citenamefont {{Sasha Weston}}}]{ElwoodAgasid2018}%
  \BibitemOpen
  \bibfield  {author} {\bibinfo {author} {\bibnamefont {{Elwood Agasid}}},
  \bibinfo {author} {\bibnamefont {{Roland Burton}}}, \bibinfo {author}
  {\bibnamefont {{Roberto Carlino}}}, \bibinfo {author} {\bibnamefont {{Gregory
  Defouw}}}, \bibinfo {author} {\bibnamefont {{Andres Dono Perez}}}, \bibinfo
  {author} {\bibnamefont {{Arif G{\"{o}}ktuğ Karacalıoğlu}}}, \bibinfo
  {author} {\bibnamefont {{Bemjamin Klamm}}}, \bibinfo {author} {\bibnamefont
  {{Abraham Rademacher}}}, \bibinfo {author} {\bibnamefont {{James
  Schalkwyck}}}, \bibinfo {author} {\bibnamefont {{Rogan Shimmin}}}, \bibinfo
  {author} {\bibnamefont {{Julia Tilles}}}, \ and\ \bibinfo {author}
  {\bibnamefont {{Sasha Weston}}},\ }\href {http://www.sti.nasa.gov} {\emph
  {\bibinfo {title} {{State of the Art Small Spacecraft Technology}}}},\
  \bibinfo {type} {Tech. Rep.}\ (\bibinfo  {institution} {NASA},\ \bibinfo
  {address} {Hanover},\ \bibinfo {year} {2018})\BibitemShut {NoStop}%
\bibitem [{\citenamefont {Bosanac}\ \emph {et~al.}(2018)\citenamefont
  {Bosanac}, \citenamefont {Cox}, \citenamefont {Howell},\ and\ \citenamefont
  {Folta}}]{BosanacNatasha2018Tdfa}%
  \BibitemOpen
  \bibfield  {author} {\bibinfo {author} {\bibfnamefont {N.}~\bibnamefont
  {Bosanac}}, \bibinfo {author} {\bibfnamefont {A.~D.}\ \bibnamefont {Cox}},
  \bibinfo {author} {\bibfnamefont {K.~C.}\ \bibnamefont {Howell}}, \ and\
  \bibinfo {author} {\bibfnamefont {D.~C.}\ \bibnamefont {Folta}},\ }\href@noop
  {} {\bibfield  {journal} {\bibinfo  {journal} {Acta astronautica}\ }\textbf
  {\bibinfo {volume} {144}},\ \bibinfo {pages} {283} (\bibinfo {year}
  {2018})}\BibitemShut {NoStop}%
\bibitem [{\citenamefont {Agasid}\ \emph {et~al.}(2021)\citenamefont {Agasid},
  \citenamefont {Hunter},\ and\ \citenamefont
  {Cheetham}}]{AgasidElwood2021CAPS}%
  \BibitemOpen
  \bibfield  {author} {\bibinfo {author} {\bibfnamefont {E.}~\bibnamefont
  {Agasid}}, \bibinfo {author} {\bibfnamefont {R.}~\bibnamefont {Hunter}}, \
  and\ \bibinfo {author} {\bibfnamefont {B.}~\bibnamefont {Cheetham}}\
  }(\bibinfo {address} {Ames Research Center},\ \bibinfo {year}
  {2021})\BibitemShut {NoStop}%
\bibitem [{\citenamefont {Cervone}\ \emph {et~al.}(2022)\citenamefont
  {Cervone}, \citenamefont {Topputo}, \citenamefont {Speretta}, \citenamefont
  {Menicucci}, \citenamefont {Turan}, \citenamefont {{Di Lizia}}, \citenamefont
  {Massari}, \citenamefont {Franzese}, \citenamefont {Giordano}, \citenamefont
  {Merisio}, \citenamefont {Labate}, \citenamefont {Pilato}, \citenamefont
  {Costa}, \citenamefont {Bertels}, \citenamefont {Thorvaldsen}, \citenamefont
  {Kukharenka}, \citenamefont {Vennekens},\ and\ \citenamefont
  {Walker}}]{CERVONE2022309}%
  \BibitemOpen
  \bibfield  {author} {\bibinfo {author} {\bibfnamefont {A.}~\bibnamefont
  {Cervone}}, \bibinfo {author} {\bibfnamefont {F.}~\bibnamefont {Topputo}},
  \bibinfo {author} {\bibfnamefont {S.}~\bibnamefont {Speretta}}, \bibinfo
  {author} {\bibfnamefont {A.}~\bibnamefont {Menicucci}}, \bibinfo {author}
  {\bibfnamefont {E.}~\bibnamefont {Turan}}, \bibinfo {author} {\bibfnamefont
  {P.}~\bibnamefont {{Di Lizia}}}, \bibinfo {author} {\bibfnamefont
  {M.}~\bibnamefont {Massari}}, \bibinfo {author} {\bibfnamefont
  {V.}~\bibnamefont {Franzese}}, \bibinfo {author} {\bibfnamefont
  {C.}~\bibnamefont {Giordano}}, \bibinfo {author} {\bibfnamefont
  {G.}~\bibnamefont {Merisio}}, \bibinfo {author} {\bibfnamefont
  {D.}~\bibnamefont {Labate}}, \bibinfo {author} {\bibfnamefont
  {G.}~\bibnamefont {Pilato}}, \bibinfo {author} {\bibfnamefont
  {E.}~\bibnamefont {Costa}}, \bibinfo {author} {\bibfnamefont
  {E.}~\bibnamefont {Bertels}}, \bibinfo {author} {\bibfnamefont
  {A.}~\bibnamefont {Thorvaldsen}}, \bibinfo {author} {\bibfnamefont
  {A.}~\bibnamefont {Kukharenka}}, \bibinfo {author} {\bibfnamefont
  {J.}~\bibnamefont {Vennekens}}, \ and\ \bibinfo {author} {\bibfnamefont
  {R.}~\bibnamefont {Walker}},\ }\href {\doibase
  https://doi.org/10.1016/j.actaastro.2022.03.032} {\bibfield  {journal}
  {\bibinfo  {journal} {Acta Astronautica}\ }\textbf {\bibinfo {volume}
  {195}},\ \bibinfo {pages} {309} (\bibinfo {year} {2022})}\BibitemShut
  {NoStop}%
\bibitem [{\citenamefont {Kruzelecky}\ \emph {et~al.}(2022)\citenamefont
  {Kruzelecky}, \citenamefont {Murzionak}, \citenamefont {Peng}, \citenamefont
  {Sinclair}, \citenamefont {Corriveau}, \citenamefont {Cloutis}, \citenamefont
  {Parkinson}, \citenamefont {Dagdick}, \citenamefont {St-Amour}, \citenamefont
  {Gao}, \citenamefont {Bridges}, \citenamefont {Baresi}, \citenamefont
  {Fabris}, \citenamefont {Rowe}, \citenamefont {Silva}, \citenamefont {Rosa},
  \citenamefont {Gameiro}, \citenamefont {Walker},\ and\ \citenamefont
  {Vennekens}}]{KruzeleckyRoman2022LVaM}%
  \BibitemOpen
  \bibfield  {author} {\bibinfo {author} {\bibfnamefont {R.}~\bibnamefont
  {Kruzelecky}}, \bibinfo {author} {\bibfnamefont {P.}~\bibnamefont
  {Murzionak}}, \bibinfo {author} {\bibfnamefont {Q.-Y.}\ \bibnamefont {Peng}},
  \bibinfo {author} {\bibfnamefont {I.}~\bibnamefont {Sinclair}}, \bibinfo
  {author} {\bibfnamefont {M.}~\bibnamefont {Corriveau}}, \bibinfo {author}
  {\bibfnamefont {E.}~\bibnamefont {Cloutis}}, \bibinfo {author} {\bibfnamefont
  {A.}~\bibnamefont {Parkinson}}, \bibinfo {author} {\bibfnamefont
  {B.}~\bibnamefont {Dagdick}}, \bibinfo {author} {\bibfnamefont
  {A.}~\bibnamefont {St-Amour}}, \bibinfo {author} {\bibfnamefont
  {Y.}~\bibnamefont {Gao}}, \bibinfo {author} {\bibfnamefont {C.}~\bibnamefont
  {Bridges}}, \bibinfo {author} {\bibfnamefont {N.}~\bibnamefont {Baresi}},
  \bibinfo {author} {\bibfnamefont {A.~L.}\ \bibnamefont {Fabris}}, \bibinfo
  {author} {\bibfnamefont {S.}~\bibnamefont {Rowe}}, \bibinfo {author}
  {\bibfnamefont {N.}~\bibnamefont {Silva}}, \bibinfo {author} {\bibfnamefont
  {P.}~\bibnamefont {Rosa}}, \bibinfo {author} {\bibfnamefont {M.}~\bibnamefont
  {Gameiro}}, \bibinfo {author} {\bibfnamefont {R.}~\bibnamefont {Walker}}, \
  and\ \bibinfo {author} {\bibfnamefont {J.}~\bibnamefont {Vennekens}}\
  }(\bibinfo  {publisher} {SPIE},\ \bibinfo {year} {2022})\ pp.\ \bibinfo
  {pages} {122360A--122360A--17}\BibitemShut {NoStop}%
\bibitem [{\citenamefont {Edelson}\ \emph {et~al.}(1979)\citenamefont
  {Edelson}, \citenamefont {Madsen}, \citenamefont {Davis},\ and\ \citenamefont
  {Garrison}}]{EdelsonRE1979VTTB}%
  \BibitemOpen
  \bibfield  {author} {\bibinfo {author} {\bibfnamefont {R.~E.}\ \bibnamefont
  {Edelson}}, \bibinfo {author} {\bibfnamefont {B.~D.}\ \bibnamefont {Madsen}},
  \bibinfo {author} {\bibfnamefont {E.~K.}\ \bibnamefont {Davis}}, \ and\
  \bibinfo {author} {\bibfnamefont {G.~W.}\ \bibnamefont {Garrison}},\
  }\href@noop {} {\bibfield  {journal} {\bibinfo  {journal} {Science (American
  Association for the Advancement of Science)}\ }\textbf {\bibinfo {volume}
  {204}},\ \bibinfo {pages} {913} (\bibinfo {year} {1979})}\BibitemShut
  {NoStop}%
\bibitem [{alm(2019)}]{alma991000261722601591}%
  \BibitemOpen
  \href@noop {} {\emph {\bibinfo {title} {Handbook of Satellite
  Applications}}}\ (\bibinfo  {publisher} {Springer New York},\ \bibinfo
  {address} {New York, NY},\ \bibinfo {year} {2019})\BibitemShut {NoStop}%
\bibitem [{\citenamefont {Friis}(1946)}]{1697062}%
  \BibitemOpen
  \bibfield  {author} {\bibinfo {author} {\bibfnamefont {H.}~\bibnamefont
  {Friis}},\ }\href {\doibase 10.1109/JRPROC.1946.234568} {\bibfield  {journal}
  {\bibinfo  {journal} {Proceedings of the IRE}\ }\textbf {\bibinfo {volume}
  {34}},\ \bibinfo {pages} {254} (\bibinfo {year} {1946})}\BibitemShut
  {NoStop}%
\bibitem [{\citenamefont {Madhow}(2008)}]{madhow_2008}%
  \BibitemOpen
  \bibfield  {author} {\bibinfo {author} {\bibfnamefont {U.}~\bibnamefont
  {Madhow}},\ }\enquote {\bibinfo {title} {Information-theoretic limits and
  their computation},}\ in\ \href {\doibase 10.1017/CBO9780511807046.007}
  {\emph {\bibinfo {booktitle} {Fundamentals of Digital Communication}}}\
  (\bibinfo  {publisher} {Cambridge University Press},\ \bibinfo {year}
  {2008})\ p.\ \bibinfo {pages} {252–292}\BibitemShut {NoStop}%
\bibitem [{\citenamefont {Tabor}\ and\ \citenamefont
  {Sibilia}(1963)}]{TaborW.J.1963MftT}%
  \BibitemOpen
  \bibfield  {author} {\bibinfo {author} {\bibfnamefont {W.~J.}\ \bibnamefont
  {Tabor}}\ and\ \bibinfo {author} {\bibfnamefont {J.~T.}\ \bibnamefont
  {Sibilia}},\ }\href@noop {} {\bibfield  {journal} {\bibinfo  {journal} {Bell
  System Technical Journal}\ }\textbf {\bibinfo {volume} {42}},\ \bibinfo
  {pages} {1863} (\bibinfo {year} {1963})}\BibitemShut {NoStop}%
\bibitem [{\citenamefont {Quinn}(1988)}]{QuinnR.B.1988A2ma}%
  \BibitemOpen
  \bibfield  {author} {\bibinfo {author} {\bibfnamefont {R.~B.}\ \bibnamefont
  {Quinn}},\ }\href@noop {} {\emph {\bibinfo {title} {A 2.3-GHz maser at Usuda,
  Japan, for TDRSS-orbiting VLBI experiment}}}\ (\bibinfo {address} {Legacy
  CDMS},\ \bibinfo {year} {1988})\BibitemShut {NoStop}%
\bibitem [{\citenamefont {Siegman}(1964)}]{siegman}%
  \BibitemOpen
  \bibfield  {author} {\bibinfo {author} {\bibfnamefont {A.~E.}\ \bibnamefont
  {Siegman}},\ }\href@noop {} {\emph {\bibinfo {title} {Microwave solid-state
  masers}}},\ McGraw-Hill electrical and electronic engineering series.\
  (\bibinfo  {publisher} {McGraw-Hill},\ \bibinfo {address} {New York},\
  \bibinfo {year} {1964})\BibitemShut {NoStop}%
\bibitem [{\citenamefont {Orton}(1970)}]{orton}%
  \BibitemOpen
  \bibfield  {author} {\bibinfo {author} {\bibfnamefont {J.~W. J.~W.}\
  \bibnamefont {Orton}},\ }\href@noop {} {\emph {\bibinfo {title} {The solid
  state maser}}},\ \bibinfo {edition} {[1st ed.].}\ ed.,\ Selected readings in
  physics / the Commonwealth and international library\ (\bibinfo  {publisher}
  {Pergamon Press},\ \bibinfo {address} {Oxford ;},\ \bibinfo {year}
  {1970})\BibitemShut {NoStop}%
\bibitem [{\citenamefont {Bertolotti}(2015)}]{alma991000618553101591}%
  \BibitemOpen
  \bibfield  {author} {\bibinfo {author} {\bibfnamefont {M.}~\bibnamefont
  {Bertolotti}},\ }\href@noop {} {\emph {\bibinfo {title} {Masers and lasers :
  an historical approach}}},\ \bibinfo {edition} {2nd}\ ed.\ (\bibinfo
  {publisher} {CRC Press},\ \bibinfo {address} {Boca Raton, Florida},\ \bibinfo
  {year} {2015 - 2015})\BibitemShut {NoStop}%
\bibitem [{\citenamefont {Oxborrow}\ \emph {et~al.}(2012)\citenamefont
  {Oxborrow}, \citenamefont {Breeze},\ and\ \citenamefont
  {Alford}}]{OXBORROWMark2012Rsm}%
  \BibitemOpen
  \bibfield  {author} {\bibinfo {author} {\bibfnamefont {M.}~\bibnamefont
  {Oxborrow}}, \bibinfo {author} {\bibfnamefont {J.~D.}\ \bibnamefont
  {Breeze}}, \ and\ \bibinfo {author} {\bibfnamefont {N.~M.}\ \bibnamefont
  {Alford}},\ }\href@noop {} {\bibfield  {journal} {\bibinfo  {journal} {Nature
  (London)}\ }\textbf {\bibinfo {volume} {488}},\ \bibinfo {pages} {353}
  (\bibinfo {year} {2012})}\BibitemShut {NoStop}%
\bibitem [{\citenamefont {Breeze}\ \emph {et~al.}(2015)\citenamefont {Breeze},
  \citenamefont {Tan}, \citenamefont {Richards}, \citenamefont {Sathian},
  \citenamefont {Oxborrow},\ and\ \citenamefont
  {Alford}}]{BreezeJonathan2015EmPe}%
  \BibitemOpen
  \bibfield  {author} {\bibinfo {author} {\bibfnamefont {J.}~\bibnamefont
  {Breeze}}, \bibinfo {author} {\bibfnamefont {K.-J.}\ \bibnamefont {Tan}},
  \bibinfo {author} {\bibfnamefont {B.}~\bibnamefont {Richards}}, \bibinfo
  {author} {\bibfnamefont {J.}~\bibnamefont {Sathian}}, \bibinfo {author}
  {\bibfnamefont {M.}~\bibnamefont {Oxborrow}}, \ and\ \bibinfo {author}
  {\bibfnamefont {N.~M.}\ \bibnamefont {Alford}},\ }\href@noop {} {\bibfield
  {journal} {\bibinfo  {journal} {Nature communications}\ }\textbf {\bibinfo
  {volume} {6}},\ \bibinfo {pages} {6215} (\bibinfo {year} {2015})}\BibitemShut
  {NoStop}%
\bibitem [{\citenamefont {Wu}\ \emph {et~al.}(2020)\citenamefont {Wu},
  \citenamefont {Xie}, \citenamefont {Ng}, \citenamefont {Mehanna},
  \citenamefont {Li}, \citenamefont {Attwood},\ and\ \citenamefont
  {Oxborrow}}]{alma991000404984801591}%
  \BibitemOpen
  \bibfield  {author} {\bibinfo {author} {\bibfnamefont {H.}~\bibnamefont
  {Wu}}, \bibinfo {author} {\bibfnamefont {X.}~\bibnamefont {Xie}}, \bibinfo
  {author} {\bibfnamefont {W.}~\bibnamefont {Ng}}, \bibinfo {author}
  {\bibfnamefont {S.}~\bibnamefont {Mehanna}}, \bibinfo {author} {\bibfnamefont
  {Y.}~\bibnamefont {Li}}, \bibinfo {author} {\bibfnamefont {M.}~\bibnamefont
  {Attwood}}, \ and\ \bibinfo {author} {\bibfnamefont {M.}~\bibnamefont
  {Oxborrow}},\ }\href@noop {} {\bibfield  {journal} {\bibinfo  {journal}
  {Phys. Rev. Applied}\ }\textbf {\bibinfo {volume} {14}} (\bibinfo {year}
  {2020})}\BibitemShut {NoStop}%
\bibitem [{\citenamefont {Ng}\ \emph {et~al.}(2022)\citenamefont {Ng},
  \citenamefont {Xu}, \citenamefont {Attwood}, \citenamefont {Wu},
  \citenamefont {Meng}, \citenamefont {Chen},\ and\ \citenamefont
  {Oxborrow}}]{https://doi.org/10.48550/arxiv.2211.06176}%
  \BibitemOpen
  \bibfield  {author} {\bibinfo {author} {\bibfnamefont {W.}~\bibnamefont
  {Ng}}, \bibinfo {author} {\bibfnamefont {X.}~\bibnamefont {Xu}}, \bibinfo
  {author} {\bibfnamefont {M.}~\bibnamefont {Attwood}}, \bibinfo {author}
  {\bibfnamefont {H.}~\bibnamefont {Wu}}, \bibinfo {author} {\bibfnamefont
  {Z.}~\bibnamefont {Meng}}, \bibinfo {author} {\bibfnamefont {X.}~\bibnamefont
  {Chen}}, \ and\ \bibinfo {author} {\bibfnamefont {M.}~\bibnamefont
  {Oxborrow}},\ }\href {\doibase 10.48550/ARXIV.2211.06176} {\enquote {\bibinfo
  {title} {{Move aside pentacene: Diazapentacene doped para-terphenyl as a
  zero-field room-temperature maser with strong coupling for cavity quantum
  electrodynamics}},}\ } (\bibinfo {year} {2022})\BibitemShut {NoStop}%
\bibitem [{\citenamefont {Moro}\ \emph {et~al.}(2022)\citenamefont {Moro},
  \citenamefont {Moret}, \citenamefont {Ghirri}, \citenamefont {del
  {\'{A}}guila}, \citenamefont {Kubozono}, \citenamefont {Beverina},\ and\
  \citenamefont {Cassinese}}]{MoroFabrizio2022Rodm}%
  \BibitemOpen
  \bibfield  {author} {\bibinfo {author} {\bibfnamefont {F.}~\bibnamefont
  {Moro}}, \bibinfo {author} {\bibfnamefont {M.}~\bibnamefont {Moret}},
  \bibinfo {author} {\bibfnamefont {A.}~\bibnamefont {Ghirri}}, \bibinfo
  {author} {\bibfnamefont {A.}~\bibnamefont {del {\'{A}}guila}}, \bibinfo
  {author} {\bibfnamefont {Y.}~\bibnamefont {Kubozono}}, \bibinfo {author}
  {\bibfnamefont {L.}~\bibnamefont {Beverina}}, \ and\ \bibinfo {author}
  {\bibfnamefont {A.}~\bibnamefont {Cassinese}},\ }\href@noop {} {\bibfield
  {journal} {\bibinfo  {journal} {Journal of materials research}\ }\textbf
  {\bibinfo {volume} {37}},\ \bibinfo {pages} {1269} (\bibinfo {year}
  {2022})}\BibitemShut {NoStop}%
\bibitem [{\citenamefont {Ng}\ \emph {et~al.}(2021)\citenamefont {Ng},
  \citenamefont {Zhang}, \citenamefont {Wu}, \citenamefont {Nevjestic},
  \citenamefont {White},\ and\ \citenamefont {Oxborrow}}]{NgWern2021EtTS}%
  \BibitemOpen
  \bibfield  {author} {\bibinfo {author} {\bibfnamefont {W.}~\bibnamefont
  {Ng}}, \bibinfo {author} {\bibfnamefont {S.}~\bibnamefont {Zhang}}, \bibinfo
  {author} {\bibfnamefont {H.}~\bibnamefont {Wu}}, \bibinfo {author}
  {\bibfnamefont {I.}~\bibnamefont {Nevjestic}}, \bibinfo {author}
  {\bibfnamefont {A.~J.~P.}\ \bibnamefont {White}}, \ and\ \bibinfo {author}
  {\bibfnamefont {M.}~\bibnamefont {Oxborrow}},\ }\href@noop {} {\bibfield
  {journal} {\bibinfo  {journal} {Journal of physical chemistry. C}\ }\textbf
  {\bibinfo {volume} {125}},\ \bibinfo {pages} {14718} (\bibinfo {year}
  {2021})}\BibitemShut {NoStop}%
\bibitem [{\citenamefont {Breeze}\ \emph {et~al.}(2018)\citenamefont {Breeze},
  \citenamefont {Salvadori}, \citenamefont {Sathian}, \citenamefont {Alford},\
  and\ \citenamefont {Kay}}]{BreezeJonathanD2018Crdm}%
  \BibitemOpen
  \bibfield  {author} {\bibinfo {author} {\bibfnamefont {J.~D.}\ \bibnamefont
  {Breeze}}, \bibinfo {author} {\bibfnamefont {E.}~\bibnamefont {Salvadori}},
  \bibinfo {author} {\bibfnamefont {J.}~\bibnamefont {Sathian}}, \bibinfo
  {author} {\bibfnamefont {N.~M.}\ \bibnamefont {Alford}}, \ and\ \bibinfo
  {author} {\bibfnamefont {C.~W.~M.}\ \bibnamefont {Kay}},\ }\href@noop {}
  {\bibfield  {journal} {\bibinfo  {journal} {Nature (London)}\ }\textbf
  {\bibinfo {volume} {555}},\ \bibinfo {pages} {493} (\bibinfo {year}
  {2018})}\BibitemShut {NoStop}%
\bibitem [{\citenamefont {Ziemkiewicz}\ and\ \citenamefont
  {Zieli\'{n}ska-Raczy\'{n}ska}(2018)}]{Ziemkiewicz:18}%
  \BibitemOpen
  \bibfield  {author} {\bibinfo {author} {\bibfnamefont {D.}~\bibnamefont
  {Ziemkiewicz}}\ and\ \bibinfo {author} {\bibfnamefont {S.}~\bibnamefont
  {Zieli\'{n}ska-Raczy\'{n}ska}},\ }\href {\doibase 10.1364/OL.43.003742}
  {\bibfield  {journal} {\bibinfo  {journal} {Opt. Lett.}\ }\textbf {\bibinfo
  {volume} {43}},\ \bibinfo {pages} {3742} (\bibinfo {year}
  {2018})}\BibitemShut {NoStop}%
\bibitem [{\citenamefont {Ziemkiewicz}\ and\ \citenamefont
  {Zieli\'{n}ska-Raczy\'{n}ska}(2019)}]{Ziemkiewicz:19}%
  \BibitemOpen
  \bibfield  {author} {\bibinfo {author} {\bibfnamefont {D.}~\bibnamefont
  {Ziemkiewicz}}\ and\ \bibinfo {author} {\bibfnamefont {S.}~\bibnamefont
  {Zieli\'{n}ska-Raczy\'{n}ska}},\ }\href {\doibase 10.1364/OE.27.016983}
  {\bibfield  {journal} {\bibinfo  {journal} {Opt. Express}\ }\textbf {\bibinfo
  {volume} {27}},\ \bibinfo {pages} {16983} (\bibinfo {year}
  {2019})}\BibitemShut {NoStop}%
\bibitem [{\citenamefont {Gottscholl}\ \emph {et~al.}(2022)\citenamefont
  {Gottscholl}, \citenamefont {Wagenhöfer}, \citenamefont {Klimmer},
  \citenamefont {Scherbel}, \citenamefont {Kasper}, \citenamefont {Baianov},
  \citenamefont {Astakhov}, \citenamefont {Dyakonov},\ and\ \citenamefont
  {Sperlich}}]{Gottscholl_2022}%
  \BibitemOpen
  \bibfield  {author} {\bibinfo {author} {\bibfnamefont {A.}~\bibnamefont
  {Gottscholl}}, \bibinfo {author} {\bibfnamefont {M.}~\bibnamefont
  {Wagenhöfer}}, \bibinfo {author} {\bibfnamefont {M.}~\bibnamefont
  {Klimmer}}, \bibinfo {author} {\bibfnamefont {S.}~\bibnamefont {Scherbel}},
  \bibinfo {author} {\bibfnamefont {C.}~\bibnamefont {Kasper}}, \bibinfo
  {author} {\bibfnamefont {V.}~\bibnamefont {Baianov}}, \bibinfo {author}
  {\bibfnamefont {G.~V.}\ \bibnamefont {Astakhov}}, \bibinfo {author}
  {\bibfnamefont {V.}~\bibnamefont {Dyakonov}}, \ and\ \bibinfo {author}
  {\bibfnamefont {A.}~\bibnamefont {Sperlich}},\ }\href {\doibase
  10.3389/fphot.2022.886354} {\bibfield  {journal} {\bibinfo  {journal}
  {Frontiers in Photonics}\ }\textbf {\bibinfo {volume} {3}} (\bibinfo {year}
  {2022}),\ 10.3389/fphot.2022.886354}\BibitemShut {NoStop}%
\bibitem [{NAS()}]{NASADist}%
  \BibitemOpen
  \href@noop {} {\enquote {\bibinfo {title} {Nasa solar system exploration},}\
  }\bibinfo {howpublished} {\url{https://solarsystem.nasa.gov/}},\ \bibinfo
  {note} {accessed: 2023-02-20}\BibitemShut {NoStop}%
\end{thebibliography}%


\end{document}